\documentclass[aps,prl,twocolumn,superscriptaddress,showpacs]{revtex4}
\usepackage{graphicx}
\draft

\begin{document}

\title{Quantum size effects in thermodynamic superconducting properties of
ultra thin films}

\author{Bin Chen}

\affiliation{Department of Physics, Oklahoma State University,
Stillwater, Oklahoma 74078, USA}

\affiliation{Department of Physics, Hangzhou Teachers College,
Hangzhou 310036, People's Republic of China}

\author{Zhenyue Zhu}

\affiliation{Department of Physics, Oklahoma State University,
Stillwater, Oklahoma 74078, USA}

\author{X. C. Xie}
\affiliation{Department of Physics, Oklahoma State University,
Stillwater, Oklahoma 74078, USA}

\affiliation{Beijing National Laboratory for Condensed Matter
Physics and Institute of Physics, The Chinese Academy of Sciences,
Beijing 100080, People's Republic of China}


\begin{abstract}
By using multi-bands BCS theory, we have calculated the
superconductivity energy gap and the critical temperature of a
thin-film metallic superconductor. The thermodynamic
superconducting characteristics such as critical magnetic field,
specific heat, as well as the tunneling conductance are
investigated for varying film thickness and temperature. We find
the oscillation of thermodynamic superconducting properties with
the film thickness, including the thermodynamic critical field
$H_c$, the specific heat of normal, superconducting state, and
position of the differential tunneling conductance peak. The two
universal constants the nth sub-band energy gap $\Delta_n$ at
temperature $T=0K$ over $k_BT_c$, and the specific heat jump at
$T_c$ over normal state specific heat at $T_c$ are independent of
the film thickness. Their values are the same as in the bulk
superconductor.
\end{abstract}


\maketitle



Quantum size effects (QSE) in superconductors have been a very
attractive topic after some theoretical investigations  of the
superconducting transition temperature $T_c$ and energy gap for thin
films.\cite{1} Some theoretical \cite{2,3,4} and experimental
\cite{5,6,7,8} works have reported that several quantities were
modulated by QSE, including electronic structure, electron-phonon
interaction, resistivity, Hall conductivity, work function, surface
energy and superconductivity critical temperature. In a metal film
on a semiconductor substrate, the conduction electrons are confined
by the vacuum on one side and the metal-semiconductor interface on
the others.\cite{9} QSE and the simple stability of metal thin films
on a supporting substrate have been discussed in several
papers.\cite{11,12,13}

The report by Guo \textit{et al.} is the first definitive and
quantitative demonstration of $T_c$ oscillation with Pb film
thickness as well as the normal state conductance.\cite{11} Using
atomically uniform film of lead with exactly known numbers of atomic
layers deposited on a silicon $(111)$ surface, Guo \textit{et al.}
observed oscillations in $T_c$ that correlated well with the
confined electronic structure and recently they demonstrated that
film thickness can indeed affect superconductivity behaviors. It has
been found that when the thickness of a film is reduced to the
nanometer scale, the film's surface and interface will confine the
motion of the electrons, leading to the formation of discrete
electronic states known as quantum well states (QWS).\cite{13} QWS
have already been observed in thin metallic films.\cite{10} And they
change the overall electronic structure of the thin film. At a tiny
thickness, physical properties are thus expected to vary
dramatically with the thickness. Very recently, Daejin \textit{et
al.} using a scanning tunneling spectroscopy study to show that both
energy gap and transition temperature exhibit persistent oscillation
without any suppression at ultra thin Pb films (5-18ML).\cite{14}

Quantum oscillations can be understood by considering QSE in these
systems.\cite{13} The period is fixed for each system and equals one
half of the bulk Fermi wavelength which is related to the average
electron density and crystal structure.\cite{12} Although lots
physical properties modulated by QSE in thin films have been
revealed, the thermodynamic superconducting characteristics such as
the specific heat, the thermodynamic critical field, tunneling
conductance as the function of thickness and other parameters have
not been reported. In this paper, by using the multi-band
superconductivity theory which improves previous theory about QSE in
superconductor film, we present the results of thermodynamic
superconducting properties of ultra-thin metallic films.

From theoretical point of view,\cite{1} the resonance and the strong
thickness dependence of $T_c$ are the characteristic features of a
thin film superconductivity. The quantization of the transverse
motion of the electron in the film leads to an increase of $T_c$
with decreasing film thickness, arising essentially from an enhanced
effective BCS pairing interaction.\cite{15} In the previous
calculations,\cite{1} the phonon modes were assumed to be the same
as in bulk material and only the one dimensional quantum confinement
effect of electrons were considered. Reference 16 considered the
phonon dispersion in a thin film undergoing substantial modification
compared with the bulk system with quantization of phonon spectrum.
They found the resonant shape of superconductor transition
temperature $T_c$ arises from both electronic and phonon
confinements. Because of the quantization of both electron and
phonon energies, the effective electron-electron interaction
modified by the quantized phonon is different from the interaction
arising from the bulk phonon.\cite{17} They found some fine
structures in the energy gap and critical temperature by comparing
with the bulk phonon model.\cite{16} However experimentally, no
evidence of such fine structure from phonon confinement was
found.\cite{11,14} Thus, we will evaluate the thermodynamic
characteristics by the same multi-band BCS theory, but neglect the
phonon quantization effects.

We will start our model by taking into account of a realistic
boundary condition. The thin film is confined in the x direction
with geometric thickness Nt. N is the thin film layer and t is the
average inter-layer distance. Because of the finite potential
barriers at the interface of thin film with Si substrates and
vacuum, we will expand the film boundaries slightly to allow certain
amount of charge spillage at two interfaces.\cite{18,19} For
simplicity, we take charge spillage distances as the same at two
interfaces. So the physical thickness is $d=Nt+2\Delta_0$.

The multi-band BCS Hamiltonian of the system is given by\cite{16}
\begin{eqnarray}
H&=&\sum_{nk\sigma}\xi_n(k)c^{\dag}_{kn\sigma}c_{kn\sigma} \nonumber\\
&+&\sum_{kk'\sigma}\sum_{nm}V^{nm}_{kk'}c^{\dag}_{k'm\sigma}
c^{\dag}_{-k'm-\sigma}c_{kn\sigma}c_{-kn-\sigma},
\end{eqnarray}
where $c^{\dag}_{kn\sigma}$ is the electron operator in the nth
sub-band with spin $\sigma$, $\xi_n(k)=\epsilon_n(k)-\mu$ the
electron energy in the sub-band n measured from the chemical
potential $\mu$, and $V^{nm}_{kk'}$ the attractive interaction
between nth sub-band and mth sub-band is given by\cite{18}
\begin{eqnarray}
V^{nm}_{kk'}=-\frac{J}{L^2}[\frac{1}{d}(1+\frac{\delta_{nm}}{2})-
\frac{b}{d^2}],
\end{eqnarray}
Only if $|\xi_n(k)|$ and $|\xi_n(k')| < \hbar\omega_D$. J is the
interaction constant and L is the periodic distance in the y and z
directions. $\hbar\omega_D$ is the Debye energy. The constant b
comes from the integral cutoff.

In the weak-coupling approximation, we have the superconducting
energy gap for the nth sub-band given by the gap function using the
mean field method
\begin{equation}
\Delta_{nk}(T)=-\sum_{m,k'}\frac{\Delta_{mk'}(T)}{2E_{mk'}}V^{nm}_{kk'}
\tanh(\frac{\beta E_{m,k'}}{2}),
\end{equation}
where $E_{mk}=(\xi^2_{mk}+\Delta_{mk}^2)^{1/2}$.
\begin{figure}[ht]
\begin{center}
\includegraphics[height=8.5cm,angle=270]{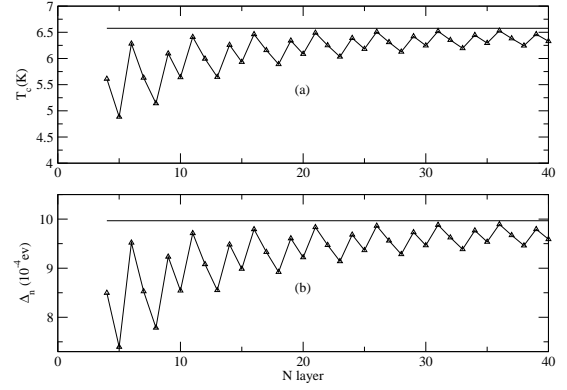}
\end{center}
\caption{(a) The critical temperature and (b) the superconducting
energy gap of the nth occupied subband versus superconducting film
layers N, compared with bulk values (straight line). Here we take
Fermi wave length $k_f=1.57{\AA}^{-1}$, interaction constant
$k=1.5{\AA}^{-1}$, Debye energy $\hbar \omega _D =0.01ev$, with
other parameters as $b=1{\AA}$, $t=2.8{\AA}$ and
$\Delta_0=0.7{\AA}$. All the following graphs are with the same
parameters. In our models, the energy gap is independent of each
subband.}
\end{figure}

The off-diagonal terms in the summation reflect the possibility of
transition of the electron pair from one sub-band into another as
the result of interaction with the confined phonons. Integration
over k gives the gap function of the nth sub-band at T. After some
derivation, we arrive at the gap equation at zero temperature
\begin{equation}
\Delta=\hbar \omega_{D}
\sinh^{-1}(\frac{Ka}{\gamma(1-b/a)+\frac{1}{2}}),
\end{equation}
The critical transition temperature is
\begin{equation}
k_{B}T_c=1.14\hbar\omega_{D}\exp(-\frac{Ka}{\gamma(1-b/a)+\frac{1}{2}}).
\end{equation}
where we assume the integral range of $k'$ in Eq. (3) are the same
for each subband. K is a special interaction constant given by
$K=\pi\hbar^2/mJ$ and $\gamma$ is the highest occupied sub-band,
given by the integer part of $dk_f/\pi$. The Fermi level only shifts
slightly within $1\%$ of its bulk value at very thin
thickness.\cite{19} So in our numerical calculation, we choose the
Fermi wave vector as a constant, independent of film thickness.
These results are similar as in Ref. 18. The bulk values of energy
gap and transition temperature are given as:\cite{18}
\begin{eqnarray}
\Delta^{bulk}&=&\hbar \omega_{D} \sinh^{-1}(\frac{K\pi}{k_f}),\\
k_{B}T_c^{bulk}&=&1.14\hbar\omega_{D}\exp(-\frac{K\pi}{k_f}).
\end{eqnarray}

In Fig. 1, we compare the results of energy gap at $T=0 K$ and the
superconducting transition temperature $T_c$ with those of bulk
values. We could see that the gap and $T_c$ all oscillate with a
period of 5 atomic layers. This can be understood by considering
that the ratio of half Fermi wave length versus atomic layer
distance is roughly $2/3$. If the layer thickness changes
continuously, the oscillation period is known to be $\lambda_f/2$.
The modlulating of these two quantities will lead the oscillation
pattern differ by 5ML. We also notice that the oscillation amplitude
will decay with increasing of the film thickness. But all these
oscillation values are smaller than bulk values. This is in accord
with several experimental results.\cite{11,14} And the most
interesting thing is that the energy gap $\Delta_n$ oscillate the
same way as $T_c$, which leads to the universal constant
$\Delta_n/k_{B}T_c=1.76$. This constant is the same as the bulk
values. Similar conclusions have been reached by experiments in
ultra thin Pb films,\cite{14} but a higher universal constant of 4.4
is obtained. Because Pb has a very strong electron phonon coupling,
the weak coupling BCS theory used here is not quite applicable.

The Gibbs free energy, the thermodynamic critical field $H_c$, the
entropy and the specific heat per volume in a superconducting film
are given, respectively  by
\begin{equation}
F=F_0+\frac{k_BT}{V}\sum_{mk}\ln[f_{mk}(1-f_{mk})],
\end{equation}
\begin{equation}
-\frac{1}{8\pi}H_c^2=\delta F(T)=-\frac{k_BT}{V}\sum_{mk}
\ln[\frac{1+\cosh(\beta E_{mk})}{1+\cosh(\beta \xi_{mk})}],
\end{equation}
\begin{figure}[ht]
\begin{center}
\includegraphics[height=8.5cm,angle=270]{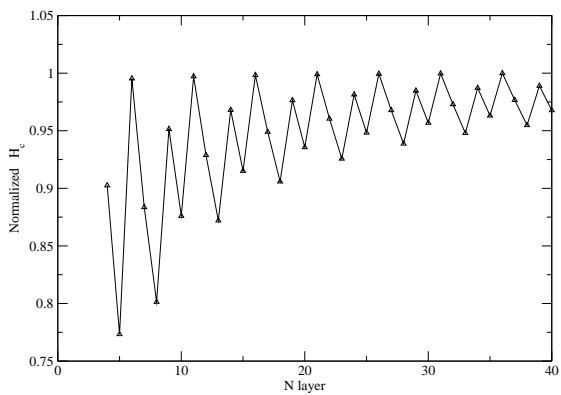}
\end{center}
\caption{The thermodynamic critical field $H_c$ in the unit of bulk
value (normalized $H_c$) as a function of film layers N in the unit
of bulk values at the reduced temperature $T/T_c=0.1$.}
\end{figure}

\begin{equation}
S=-\frac{2}{V}\sum_{mk}[(1-f_{mk})\ln(1-f_{mk})+f_{mk}\ln f_{mk}],
\end{equation}
\begin{equation}
C_v=\frac{2\beta^2 k_B }{V}\sum_{mk} f_{km}(1-f_{km})
(E_k^2+\frac12\beta\frac{\partial \Delta_m^2}{\partial\beta}),
\end{equation}
where $f_{km}=[\exp(\beta E_{km})+1]^{-1}$ and $\beta=1/k_BT$.
\begin{figure}[ht]
\begin{center}
\includegraphics[height=8.5cm,angle=270]{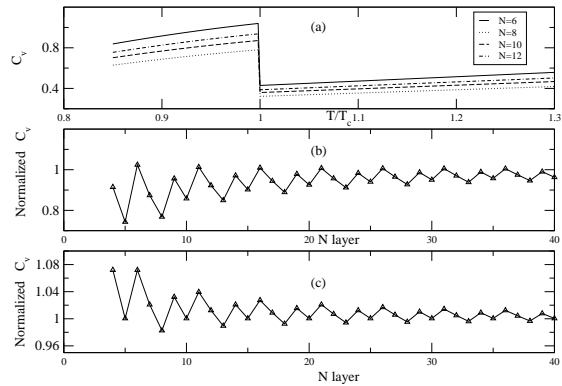}
\end{center}
\caption{(a) The variation of specific heat with reduced temperature
$T/T_c$ for different film thickness. The normalized specific heat
versus film layers N at superconducting state ($T/T_c=0.9$) and at
normal state (T=8k) is shown respectively in (b) and (c).}
\end{figure}

The thermodynamics critical field $H_c$ is also shown in Fig. 2. We
also find the similar oscillation of $H_c$ with the the film layers
N, although no experimental are available yet. The layer dependent
$H_c$ shows an oscillation amplitude of $22\%$ of the bulk situation
at 5 to 8ML and decay to the amplitude of $5\%$ of the bulk $H_c$ at
thicker films. However, QSE on the perpendicular upper critical
field $H_{c2}$ in the ultra-thin lead film has been reported whose
oscillation is out of phase with the $H_c$ oscillation.\cite{20}

The specific heat is also calculated with Eq. (11) at both
superconductor and normal state. In Fig. 3(a), we show the specific
heat as a function of reduced temperature $\lambda$
$(\lambda\equiv/T_c)$ at different films layers. We could see the
jump of specific heat at critical temperature. Meanwhile, we also
notice that although the specific heat both at normal and
superconducting state oscillate with film thickness, the jump of
specific heat at $T_c$ divided by the normal state specific heat
$[(C_s-C_n)/C_n]_{T_c}$ is another universal constant 1.42, and this
constant is the same as in bulk situation. In Fig. 3(b) and 3(c), we
can see that the specific heat at supercondcting state varies in the
same way as $T_c$ does. But its value could be slightly higher than
the bulk case. For the specific heat in a normal state, it oscillate
unsymmetrically around its bulk value. This oscillation behavior is
different from others. Hopefully, all these results could be
verified in future experiments.

The ratio of superconductor-normal tunneling differential
conductance $G_{SN}$ to normal tunneling differential conductance
$G_{NN}$ is
\begin{equation}
\frac{G_{SN}}{G_{NN}}=\int
\frac{sign(\varepsilon)}{\sqrt{\varepsilon^2-\Delta^2}}
\frac{\partial f(\varepsilon+eV)}{\partial \varepsilon} d
\varepsilon.
\end{equation}
\begin{figure}[ht]
\begin{center}
\includegraphics[height=8.5cm,angle=270]{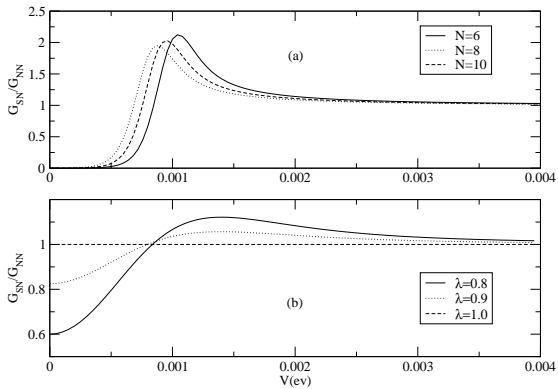}
\end{center}
\caption{$\frac{G_{SN}}{G_{NN}}$ as the function of applied voltage
V (a) at different film thickness with $T=1K$  and (b) for different
reduced temperature $\lambda$ with film layers N=7.}
\end{figure}
The tunneling differential conductance versus bias voltage
calculated at different film layers and different temperature from
Eq. (12) are also shown in Fig. 4. The peak of the conductance is
almost at $eV=\Delta_n$. Because in our model all the energy gaps of
each sub-band are assumed to be the same, we could only find one
peak in the tunneling conductance. We also find that the position of
the conductance peak oscillates with thickness. The reason for this
kind of oscillation is due to the oscillation of energy gap
$\Delta_n$ at different thicknesses. In Fig. 4(b), we can see that
the peak is depressed at high temperature, because energy gap
decreases with increasing temperature. Recently, Daejin \textit{et
al.} have already extracted energy gap from the measured conductance
spectra to determine the transition temperature.\cite{14}

In this paper, we apply multi-band BCS theory to calculate
thermodynamic quantities and find their oscillations with the film
thickness, including the thermodynamic critical field $H_c$, the
specific heat at normal and superconducting state and the position
of differential conductance peak. These oscillations are the
manifestations of the QSE. But two universal constants
$\Delta_n/k_{B}T_c=1.76$ and $[(C_s-C_n)/C_n]_{T_c}=1.42$ is
independent of the film thickness, and their values are the same as
in bulk situation. We hope that these interesting oscillations could
all be observed experimentally in the future.

We greatly acknowledge financial support from US DOE under grant No.
DE-FG02-04ER46124 and from NSF under grant No. CCF-0524673. B.C. is
supported by NSF China under grants No. 10574035 and No. 10274070.
Z.Z. also acknowledges support by a Grant-In-Aid of Research from
Sigma Xi Society.

\end{document}